\documentclass[12pt]{article}
\usepackage{epsf}
\usepackage{slashed}
\usepackage{color}
\usepackage{hyperref}
\begin{document}
\begin{titlepage}
\title{
CONNECTING GRAVITY AND STRONG INTERACTIONS}
\author{
{Piotr \.Zenczykowski }\footnote{E-mail: piotr.zenczykowski@ifj.edu.pl}\\
{\em Professor Emeritus}\\
{\em The Henryk Niewodnicza\'nski Institute of Nuclear Physics}\\
{\em Polish Academy of Sciences}\\
{\em Radzikowskiego 152,
31-342 Krak\'ow, Poland}\\
}
\maketitle
\begin{abstract}
We use experiment-supported dimensional analysis to further bolster
our arguments that crucial information on the emergence and/or 
nature of space could be
extracted from the combination of the properties of gravitational and strong interactions.\\
\end{abstract}

\vfill
\end{titlepage}

\section{INTRODUCTION}

It is believed by the majority of physicists that spacetime
emerges at the scale of Planck length, 
as hinted at by the 
 dimensional analysis based on the  
gravitational ($G$),  relativistic ($c$),
and quantum ($h$) constants. 
Yet, in \cite{Zen2018}
it was suggested that it is the properties of hadrons,
20 orders of magnitude above Planck length, that should give us important information 
on the emergence
or structure of space.
As this is an unorthodox view it requires a solid justification.
The idea, originally presented in \cite{Zen2018}, 
was discussed later in \cite{Zen2021}.
In the present paper we will reconsider the issue from the point
of  dimensional analysis combined with experimental hints.

In 1921 Einstein, presenting to the general public in simple words
his stance on the connection between matter and space, said:
{\it ``time and space disappear together with things''} \cite{Einstein}, (see also \cite{Pen1968}). Following this idea, properties of space should be connected with and probably derivable from those of matter. 
With hadronic matter constituting the bulk of the observable mass in the Universe
it seems that a rough structure of the 3D macroscopic space that surrounds us should first 
appear at the hadronic mass and distance scales. 
In fact, in \cite{Pen1974} Penrose dismissed the arguments that the hadronic distance scale is too large stressing that: {\it `` It is the proton
itself, not the spacetime point, which behaves as a discrete physical entity and which has,
at least to a considerable degree, some semblance of indivisibility."}

The arguments used in \cite{Zen2018} included 
the dimensional analysis which involved the cosmological constant $\Lambda$, while those of \cite{Zen2021} discussed an approach based on $a_M$, the MONDian bound on acceleration. 
With physics being ultimately based on experiment in this note we want 
to shed some further light on the MOND argument,  
stressing the crucial role of experimental input in choosing the relevant fundamental constants. 
\\
  
\section{DIMENSIONAL ANALYSIS}
We consider dimensional analysis as the first step that should be tried
before any more specific approach is proposed.
First, however, following Meschini \cite{Meschini}, we must recall that
dimensional analysis leads to proper conclusions only 
when one uses constants appropriate to a problem under consideration.
The choice of fundamental constants is therefore absolutely crucial, and experiment
should play a decisive role in supplying this choice.\\

Now, the Einstein equations contain not only the Newtonian 
gravitational constant $G$ but also 
the cosmological constant $\Lambda$, increasing to 4 the number of fundamental
constants to be possibly used in dimensional analysis.
Thus, there are three possible discrete choices for 
the relevant mass and distance quantum 
scales \cite{Zen2018}:\\

1. the Planck scale, based on the choice of $h$, $G$, and $c$ (with $\Lambda$ not used), which 
                   gives $m_P = \sqrt{hc/G} = 5.46 * 10^{-5} g$, 
                   and $l_P = h/(m_P c)= \sqrt{hG/c^3}    = 4.05 * 10^{-33} cm$;\\

2. the Wesson scale, based on the choice of $h$, $\Lambda$, and $c$ (with $G$ not used), which gives $m_W =h/c \sqrt{\Lambda/3} = 1.39 * 10^{-65} g$
             and $r_W = h/(m_W c) = \sqrt{3/\Lambda} = 0.9 * 10^{28} cm  \approx r_U$,
		with $r_U$ being the size of the Universe; and\\

3. the nonrelativistic / hadronic scale, 
based on the choice of $h$, $G$, and $\Lambda$ (with $c$ not used), which gives 
                       $m_H =(h^2/G * \sqrt{\Lambda/3})^{1/3}= 0.35 * 10^{-24} g$
                       (to be compared with proton mass $m_p=1.67 * 10^{-24}~g$) 
             and       $r_H = h/(m_H c) = 10^{-12} cm$.\\                  

Now,
choices \# 1 and \# 2 probe completely 
untestable regions of phase space ($10^{-33} cm$, $10^{-65} g$)
and may constitute artefacts of our quantum-theoretical speculations.
After all, can quantum theory be soundly extrapolated from the atomic or elementary
particle distance and mass scales
to such tiny distances and/or masses? Do such scales have any recognizable
physical meaning?\\

On the other hand, choice \# 3 singles out the experimentally accessible 
region of (hadronic) mass $m_H$ and distance $r_H$ scales where 
theoretical quantum description (ie. the Standard Model) 
has been corroborated  
as an excellent approximation of certain aspects of reality.
In particular, choice \# 3 leads to an agreement 
with experimentally determined hadronic scales ($m_H \approx 10^{-24}~g$, $r_H \approx 10^{-12}~cm$), 
which agreement is - in my opinion - extremely unlikely to be a mere coincidence.
The applicability of choice \# 3 and the relevance of hadrons is therefore strongly
indicated.\\

Note that $r_H/l_P = m_P/m_H = N^{1/6} = 10^{20}$ 
(where $N=c^3(h G \Lambda)=15 *10^{120}$ \cite{Zen2018})
ie. $l_P$ differs from $r_H$ by 20 orders of magnitude.
Moreover, $m_P=m_H N^{1/6}$; $m_W=m_H N^{-1/3}$, ie. 
$m_W$ and $m_P$ differ by 60 orders of magnitude
($m_P/m_W = N^{1/2} = 10^{60}$, a cube of $r_H/l_P$).
Thus, the deviations of $l_P$ and $m_W$ from the hadronic 
values of $r_H$ and $m_H$ are defined by the same combination
of constants $(N^{1/6})$.
Probably, therefore, $l_P/r_H$ and $m_P/m_W$  are of 
the same origin and may be both artefacts of the 
application of the quantum description beyond 
its range of validity. 
In fact, \cite{Bojowald 2007} argued that the typical 
distance scale relevant for the idea of space  
quantization ($l_{SQ}$) may be much larger than $l_P$.\\
 
 The fully classical 
choice 4. $G$, $c$, and $\Lambda$ (with $h$ not used) 
gives $(m_U,r_U)$, ie. the mass and size of the Universe:
       $m_U=c^2/(G) \sqrt{3/\Lambda})=2.14*10^{56} g$, 
       $r_U= \sqrt{\Lambda^{-1}/3}=0.9 *10^{28} cm$.\\
       
       In this paper an argument corroborating those used in \cite{Zen2018},
ie. selecting {\it experimentally supported}  
choice \# 3, while dismissing physically suspected choices \# 1 and \# 2 as experimentally unverified 
purely theoretical speculations, will be used to argue again  
that  interesting and probably crucial aspects of the onset of the emergence 
of space should occur and be observable at hadronic scales.

\section{THEORY'S DOMAIN}

Any physical theory provides a description of some particular aspects
 of nature and in certain ways only.
We always have idealization and approximation at work,
as only some aspects of reality are intentionally or of necessity
taken into account, while others are not\cite{Heisenberg1958}.
In  classical or quantum theories
space is treated classically as a continuum background 
both for macroscopic objects (in classical theories)
as well as for elementary particles (in the quantum field theoretical Standard Model).
The latter description constitutes therefore 
a classical-quantum hybrid \cite{Finkelstein1969}, 
which - while formally and numerically very successful - 
exhibits our lack of understanding of the connection between classical
and quantum descriptions, ie. 
between classical locality (separability) and quantum nonlocality (nonseparability), 
as well as of
the crucial role of spatial superpositions in quantum description   
and their lack in the macroscopic classical world, etc.\\

It may be argued that such conceptual clashes result 
from illegitimate carry-over of
the classical arena to quantum theory
(ie. from the hybrid nature of QT) \cite{Singh QTandST1707.01012}.
After all, classical and quantum theories 
constitute distinct descriptions of nature,
which only approximate its different aspects. 
The use of the classical conception of 
the underlying spatial continuum in QFT should be treated 
as a macroscopic approximation and idealization, 
and therefore it must have some physically set limits of applicability. 
The question then is what are these limits.
As the answer should be physically meaningful, 
this question is predominantly an experimental one.
Theory can only suggest in which direction one should go.\\

Now, if matter and space are related, the simplest speculation 
is to extend the Democritean idea 
from atoms of matter to atoms of space. 
Since in General Relativity the properties of space are subject to 
the classically determined relativistic dynamics, 
the relevant expectation 
on their spatial size seems to be provided 
by dimensional analysis involving classical constants 
describing gravity and relativity $c$  
supplemented with the quantum constant $h$.  
With the forces of Newtonian gravity being described 
by the gravitational constant $G$, it may seem
that the size of the atoms of space should indeed be 
$l_P \approx  \sqrt{hG/c^3}$ (choice 1).
Yet, this reasoning is a theoretical speculation only. 
There is no experimental hint that it has some 
correspondence with reality \cite{Meschini}.
Furthermore, it does not touch 
the really fundamental quantum-classical conceptual 
clashes mentioned above. \\

For example, one of the problems concerns
the interpretation of the quantum
prescription for measurement 
and the transition 
  from the probabilistic description
of local field theories (eg. the Standard Model) 
whose quantum aspects are characterized 
by linear superposition of 
amplitudes from different points of space 
(thus involving nonlocality and nonseparability),   
  to the classical deterministic description 
in which things are localized
and separable in space, and cannot be found 
in two places at the same time. \\

Among various interpretations of the
quantum prescription there is a class that assumes  
that quantum and classical descriptions 
constitute limiting cases of a single, 
broader and objective theory,
and that they both miss a part that actually accounts for a continuous spontaneous
localisation (objective collapses of the quantum wave function) 
that prevents superposition of objects 
in the classical world \cite{GRW}.
This broader theory is assumed to depend on 
new fundamental constants \cite{GRW},\cite{CSL}:
the radius of localization length $l_C$ 
and the rate of collapse $\lambda$.
There is an agreement among workers in the field
that the size of collapse localisation region
is of the order of $l_C = 10^{-5}~ cm$; while the
rate is estimated to be $10^{-17} s^{-1}$ \cite{GRW},\cite{CSL}
or $10^{-8}~s^{-1}$ \cite{Adler}.
Note that these numbers are theoretical only 
and that the localization radius is many orders of magnitude
larger than the Planck length:  $l_C = 10^{+28} l_P$  
(with $m_C =    10^{-28} m_P = 10^{-33}~g$ ).
Planck length $l_P$ and collapse length $l_C$ give us two very different theoretical 
guesses for the lower bound on the classical size $l_{SQ}$ of a region to 
which the approximation of the conception of 
a mathematical point is still acceptable. These are
based on different physics (GR and spontaneous localisation)
and differ enormously \cite{Bojowald 2007}. 
The first is 
experimentally unreachable, the second is nearly macroscopic
 and should be testable. Thus, various experiments are going on.

Now, General Relativity describes the action of gravity/matter 
through the properties of classical space.
This suggests that space is a consequence of the existence 
of matter and the gravitational field it generates.
Since quantum collapse is to create space \cite{Singh QTandST1707.01012}, it
should be related to gravity. Indeed, the Diosi-Penrose model 
\cite{Diosi1989},\cite{Penrose1996}, 
associates the spontaneous collapse of the wave function and 
the emergence of ordinary space to the properties of gravity.
Quite generally, with space viewed as a property of matter, 
the limits beyond which our macroscopic conception of
space is no longer applicable should concern particle 
masses or the strength of the matter-induced gravitational field (or both). 

\section {BOUNDS}

In the analysis of paper \cite{Zen2018} four possibly relevant 
fundamental constants were used: $h$, $c$, $G$, and
$\Lambda$.
These constants belong to two 
different groups: 1) the boundary constants that determine
the domains of applicability of theoretical quantum and relativistic concepts 
(such as $h < action$, $speed < c$), 
and 2) the size of other theoretical concepts 
(such as gravitational constant $G$ and probably 
the cosmological constant $\Lambda$) 
used within such domains but not determining them. In particular, as noted
by Milgrom, 
\cite{Milgrom2019}
$G$ is essentially a conversion factor 
that links the gravitational and inertial masses. 
By itself it does not set up 
a physically-defined gravitational bound which we expect
to exist by analogy with quantum and relativistic bounds. 
Thus, so far we have only two of the three
bounds expected  to set up a closed/bounded domain in the mass(momentum)-position space.\\

Indeed, as far as gravity is concerned, we have not yet specified
the relevant bounding constant. Now,
since the gravitational field is described by 
the acceleration it induces, the relevant bounds should 
probably be expressed as 
limits on the acceleration. In fact, limits on acceleration may be argued to be more fundamental
than corresponding limits on distance: after all, it is acceleration that describes the
strength of matter-induced gravitational field and its connection with the properties
of space.
By dimensional arguments we have theoretical upper bounds on acceleration
$a_P=c^2/l_P = c^2 \sqrt{c^3/hG} = \sqrt{c^7/(hG)} = 2.25 \times 10^{53}~ cm/s^2$ and
$a_C=c^2/l_C = 10^{-28} a_P = 10^{25}~cm/s^2$. However, both these numbers constitute theoretical
speculations only. We badly need experimental or observational input. Such an input
may be experimentally observed when looking for $l_C$. Yet,
it appears that there already is a different experimental candidate. It comes from astrophysics. 
Namely, astrophysical observations suggest that something strange happens 
at $a_M = 1.2 * 10^{-8}~ cm/s^2$, which seems to behave like another fundamental bounding constant. 
The relevant idea, 
driven by the observed flat shape of stellar rotation curves 
for stars on the outskirts of galaxies, was proposed 
by Milgrom 40 years ago \cite{Milgrom1983}.
As an explanation of these observations of stars
moving too fast when compared with the expectations based on the 
assumption of Newtonian gravitational forces, 
Milgrom suggested that far away from Galaxy centers 
the acceleration $a$ induced by relevant
galactic matter ($m$) instead of
dropping in the Newtonian fashion, like

\begin{equation}
\label{Newton}
a = a_N = Gm/r^2, 
\end{equation}
it starts to fall off
linearly with distance, following the formula:
\begin{equation}
\label{MOND}
a = \frac{\sqrt{Gm}}{r} \sqrt{a_M}. 
\end{equation}

 Hence the name Modified Newtonian Dynamics (MOND).
 Whatever the deep meaning of MOND,
over the years it became a strong 
competitor of the dark matter paradigm.
An extensive list of solid physical arguments in favor 
of MOND is gathered in \cite{proMOND}. 

Comparing (\ref{Newton}) and (\ref{MOND}) one finds the distance $r_M$ 
from the center of Galaxy of mass $m$ at which the transition from the Newtonian
to MONDian form occurs:

\begin{equation}
\label{MONDradius}
r_M =\sqrt{G/a_M} \sqrt{m}.
\end{equation}

Constants $h$, $c$, and $a_M$  define three bounds  imposed by 
quantum, relativistic and gravitational aspects of physical reality
together specifying classical domain in the ($m-r$) phase space:
($action > h$,
$speed < c$, and
$acceleration > a_M$).
Note that $G$ and $a_M$ enter in combination $a_M/G$ only,
linking space to matter through
\begin{equation}
a_M/G = m/r^2_M,   
\end{equation}
with
\begin{equation}
\label{discmass}
a_M/G = 0.18 ~g/cm^2.
\end{equation}

The quantum connection of matter to space is
defined by the Compton wavelength

\begin{equation}
\label{Compton}
r_Q m = h/c. 
\end{equation}

Solving  Eqs (\ref{MOND},\ref{Compton}) for $m$ by putting $r_M=r_Q=r$ 
gives

\begin{equation}
\label{massH}
m = ((h/c)^2 a_M/G)^{1/3} 
\end{equation}

ie. $m=0.2 *10^{-24} ~g \approx m_H$,

and radius 
\begin{equation}
\label{radiusH}
r = (G/a_M * h/c)^{1/3} = 10^{-12}~ cm \approx r_H. 
\end{equation}

Note that in formal limit $a_M,h,1/c\to 0$ the mass $m$ in (\ref{massH}) goes to zero:
the above estimate gives thus the lower bound for mass
built as proportional to $(acceleration * (action/velocity)^2)^{1/3}$.
Eq.(\ref{massH}) gives essentially the same estimate as choice 3 of sect. I.
This is because of the well-known coincidence  

\begin{equation}
\label{coincidence}
c^2 \sqrt{\Lambda} \approx 8.2 * a_M . 
\end{equation}

Formulas (\ref{massH}),(\ref{radiusH}) can also be obtained by simple dimensional
considerations using three fundamental {\it bounding} constants:
(relativistic $c$, quantum $h$, and gravitational $a_M/G$),
that replace the Planck choice of $c$, $h$, and $G$.
Similarly, formula (\ref{discmass}) gives an estimate of the planar 
mass density of hadrons, ie. hadronic mass/hadronic cross section

\begin{equation}
\label{G-S}
a_M/G = m_H/r^2_H \approx 10^{-24}/10^{{-12}*2}~g/cm^2 \approx 1~g/cm^2.
\end{equation}

This formula that equates a ratio of two observed general characteristics
of gravity to a ratio of two important experimental characteristics of strongly
interacting particles is to me simply stunning.
Formula (\ref{G-S}) suggests a deep physical connection between the spatial
dimensionality of the two pairs of factors appearing on its left and right sides,
namely 3D associated with $m_H$ and $1/G$, and 2D for $1/r_H^2$ and $a_M$ (with
Gauss law in 2D suggested by (\ref{MOND})).  
Consistency of the treatment of dimensions on the two sides of (\ref{G-S}) 
suggests that formula (\ref{MOND}) describes gravity in 2D.
I tend to believe that it is from the tiny accelerations of  2D nature 
that stronger accelerations,
of the 3D nature are composed. The prescription for the build up of Newtonian
gravity and 3D space seems to start from physical 2D subspaces.

\vfill

\vfill


\begin{thebibliography}{99}
\bibitem{Zen2018} P. \.Zenczykowski, Found. Sci. 24 (2019) 287. 
\bibitem{Zen2021} P. \.Zenczykowski, Mod. Phys. Lett. A 2150252 (2021); arxiv:2112.07357.
\bibitem{Einstein} https://www.nature.com/articles/d41586-018-05004-4.
\bibitem{Pen1968} R. Penrose, Structure of Spacetime, in C.M. DeWitt and J.A. Wheeler (eds)
{\it Batelle Rencontres} (New York, 1968) p.121.
\bibitem{Pen1974} R. Penrose, Twistors and Particles: an Outline, in {\it Proc. of the Conference on Quantum Theory and the Structures of Time and Space} (Feldafing, July 1974), p. 129.
\bibitem{Meschini} D. Meschini, Found. Sci. 12 (2007) 277.
\bibitem{Bojowald 2007} M. Bojowald, Quantum gravity and cosmological observations. {\it AIP Conference Proceedings}, 917, 130–137; arXiv: gr-qc/0701142.
\bibitem{Heisenberg1958} Heisenberg, W. (1958). {\it Physics and philosophy: The revolution in modern science} (p. 200). New York: Harper and Row.
\bibitem{Finkelstein1969} Finkelstein, D. R. (1969), Phys. Rev., 184, 1261–1271.
\bibitem{Singh QTandST1707.01012}
T. P. Singh, 
 Z.Naturforsch.A 73 (2018) 8, 733-739; 1707.01012. 
 T. P. Singh,
 
 
 
  
 Z.Naturforsch.A 73 (2018) 10, 923-929; 1806.01297. 
\bibitem{GRW} 
G. C. Ghirardi, A. Rimini, and T. Weber, Phys. Rev. D 34, 470 (1986).
\bibitem{CSL} G. C. Ghirardi, P. Pearle, and A. Rimini, Phys. Rev. A 42, 78 (1990).
\bibitem{Adler} S. Adler, 
J. of Phys. A: Math. and Theor. 40 (44): 13501 (2007); arXiv:quant-ph/0605072.
 \bibitem{Diosi1989} Diósi, L.,  
Phys. Rev. A. 40 (3): 1165–1174 (1989).
\bibitem{Penrose1996} Penrose, R., 
Gen. Relat. and Grav. 28 (5): 581–600. (1996)
 \bibitem{Milgrom2019} M. Milgrom, arXiv:1910.04368.
\bibitem{Milgrom1983} M. Milgrom, Astrophys. J. 270 (1983) 365.
\bibitem{proMOND} S. McGaugh, Can. J. Phys. 93 (2015) 250;
 R. Scarpa, AIP Conf. Proc. 822 (2006) 253, arXiv:astro-ph/0601478; 
M. Milgrom, arXiv:0908.3842; B. Famaey and S. McGaugh, Living Rev. Relativity 15
(2012) 10.
\end{thebibliography}
\end{document}